# Scalable Production of Photochromic Yttrium Oxyhydride Powder via Ball Milling

Elbruz Murat Baba[*], Stefano Deledda, Smagul Karazhanov[†]

Institute for Energy Technology, PO BOX 40, Kjeller, Norway

## Abstract

Yttrium oxyhydride (YHO) represents one of the most promising photochromic materials discovered in recent years, yet its practical deployment has been severely constrained by the limitations of thin-film deposition methods. Here we demonstrate the first successful synthesis of photochromic YHO powders through reactive ball milling under hydrogen atmosphere followed by controlled oxidation - a fundamentally scalable approach that overcomes the production barriers facing this technology. High-energy planetary ball milling of yttrium metal under 50 bar $H_2$ for 20 hours, followed by controlled oxidation in ultra-dry technical air ($H_2O$ < 0.5 ppm), yielded nanostructured YHO powders with sub-500 nm particle sizes. These powders exhibit robust photochromic response with ~10% reflectance modulation at 850 nm under 405 nm excitation, reversible cycling behavior, and the characteristic "memory effect" previously observed only in thin films. Powder X-ray diffraction confirms the formation of the cubic YHO phase with lattice expansion consistent with oxygen incorporation into the yttrium hydride structure. Critically, we demonstrate that YHO powders can be processed into polymer composites enabling spatially-resolved photochromic patterning - a capability essential for practical device applications. While optimization of optical contrast remains an opportunity for future work, this powder synthesis route fundamentally transforms the manufacturing of YHO-based photochromic systems, enabling mass-scale production using established industrial ball milling infrastructure. These findings establish a viable pathway toward commercial deployment of YHO in smart windows, adaptive optics, and rewritable information storage applications.

## Introduction

Yttrium oxyhydride (YHO) is an emerging compound with promising properties, particularly its photochromic behaviour, making it a strong candidate for various advanced technological applications. To date, research primarily focused on deposition of yttrium oxyhydride via reactive sputtering [1], [2], [3], [4], [5] and e-beam evaporation [6], [7] techniques, yielding photochromic thin-films. In contrast, the fabrication of photochromic yttrium oxyhydride in powder form remains largely unexplored. Materials capable of optical modulation in response to irradiance, especially solar irradiance, are in high demand across industrial sectors. Among these, photochromic materials stand out as true smart materials, exhibiting significant absorption in the visible spectrum and offering the potential to mitigate one of the most pressing challenges such as carbon emissions.

Currently, various photochromic powder materials are commercially available, including those based on metal oxides, organic compounds, and other chemical systems. However, attempts to produce photochromic powders from rare-earth metal oxyhydrides, a novel and rapidly developing class of materials, have so far been unsuccessful. One study reported the chemical synthesis of neodymium oxyhydride (NdHO) powder by combining metal hydrides with neodymium oxide [8] though no photochromic behaviour was observed. Zapp et. al. [9] explored four distinct methods for synthesising YHO powders using combinations of (a) yttria ($Y_2O_3$) and yttrium hydride ($YH_3$); (b) yttria and calcium hydride ($CaH_2$); (c) yttrium hydride and calcium oxide (CaO); and (d) yttrium oxide fluoride (YOF) and alkaline hydride (AH where A is Lithium or Sodium). Nevertheless, none of the resulting YHO powders exhibited photochromic properties.

The production of photochromic yttrium oxyhydride was previously demonstrated by Mongstad et al. [3], You et al. [10] and Montero et. Al [2]. To date, studies have reported photochromic thin films synthesized via physical vapor deposition methods, mainly reactive magnetron sputtering, then oxidising in air. This study reports, for the first time, a method for producing nanoscale photochromic yttrium oxyhydride powders suitable for mass production. This represents a significant advancement, as powder-based materials offer distinct advantages for large-scale production and diverse applications compared to thin films. The reactive ball milling and subsequent controlled oxidation process described herein provides a scalable route to obtain YHO powders with controllable particle size. The demonstration of photochromic behavior in these YHO powders, comparable to that observed in thin films [11], opens up new possibilities for utilizing this material in applications where bulk quantities are required, such as pigments, inks, or coatings. Furthermore, this method's scalability suggests potential cost-effectiveness for industrial production, making it a promising alternative for widespread implementation of photochromic technologies.

---

[*] Corresponding author. E-mail: elbruz.baba@ife.no

[†] Corresponding author. E-mail: smagul.karazhanov@ife.no

## Methodology

The synthesis of yttrium oxyhydride powder was conducted using a high-energy reactive planetary ball mill, employing hardened steel milling vials and balls under an argon (5N purity) and hydrogen (5N purity) atmosphere followed by oxidation. Yttrium metal pieces (3N purity) were loaded into the hardened-steel vial along with hardened steel balls. The lid of the milling vial was equipped with a gas inlet and valve to evacuate the inner volume of the ball milling vial or fill it with gas. The lid also housed a pressure transducer and thermometer to monitor the pressure and temperature inside the milling vial. The ball to starting material mass ratio was maintained within the range of 10:1 to 50:1. After loading the starting material and balls, the milling vial lid was tightened to create an air/gas-tight seal. The vial was then connected to a vacuum pump via the gas inlet to evacuate any air. Once a vacuum of approximately $10^{-3}$ mbar was achieved, the vial was filled with hydrogen until a pressure of up to 50 bar was attained. With the vial filled with hydrogen, it was mounted in a Fritsch P6 planetary ball mill, and the milling process was initiated. Milling was carried out at a rotational velocity between 300 and 500 rpm for up to 20 hours. After the milling process, powders were collected inside an Argon atmosphere and then transferred to a technical dry air (H2O < 0.5 ppm, O2 < 5-10%) filled glovebox for oxidation. Samples were kept inside the glovebox for oxidation more than 2 days. Optical characterization measurements were performed using an Ocean Optics QE6500 spectrometer to measure reflectance, and a RC PRO Digital Microscope was used to record and extract the red component from images capturing the cyclic change between the photodarkened and bleached states.

Powder X-ray Diffraction (PXD) was performed with a Bruker D2 Phaser in a Bragg Brentano configuration using Cu-Kα radiation (λ = 1.5406 Å). PXD measurements were carried out in air using Si low background sample holders which were rotating during acquisition.

The microstructure of photochromic yttrium oxyhydride nanoparticles were analyzed using a high-resolution field emission Scanning Electron Microscope (SEM), operated in secondary electron mode at an accelerating voltage of 5 kV.

## Results

Photochromic YHO powder synthesis investigated in this work involves two main steps: firstly, yttrium pieces were crushed through ball-milling inside a hardened-steel vial containing high pressure hydrogen, then the as-milled powder particles were oxidized under technical grade dry air to reduce the oxidation rate. The initial ball milling step is illustrated in Figure 1, which presents the evolution of hydrogen pressure and temperature inside the milling vial containing yttrium pieces over a 5-hour milling period. In the first stages of the milling process, the temperature of the vial gradually increases from room temperature up to almost 50 °C after about 1 hour of milling, primarily due to mechanical impacts and friction from the milling balls. After about 4 hours, thermal equilibrium is reached, with the vial stabilizing at 52.5 °C as heat generation and dissipation balance out. The hydrogen pressure also increases in the first minutes of milling due to the temperature increase but starts decreasing when $H_2$ is absorbed by the ball milled material. Hydrogen absorption (and the decrease of pressure) continues for about 1.5 hours of milling. After that, the $H_2$ pressure remains constant, indicating that ball milled material is saturated with hydrogen. In Figure 1, the overall change of $H_2$ pressure measured at room temperature inside the vial before and after the milling run was ΔP = 10 bar. Considering the internal volume of the vial not occupied by the balls and by the ball milled materials, this change of pressure corresponds to about 7.8 $10^2$ mol of $H_2$ being absorbed by the ball milled material.

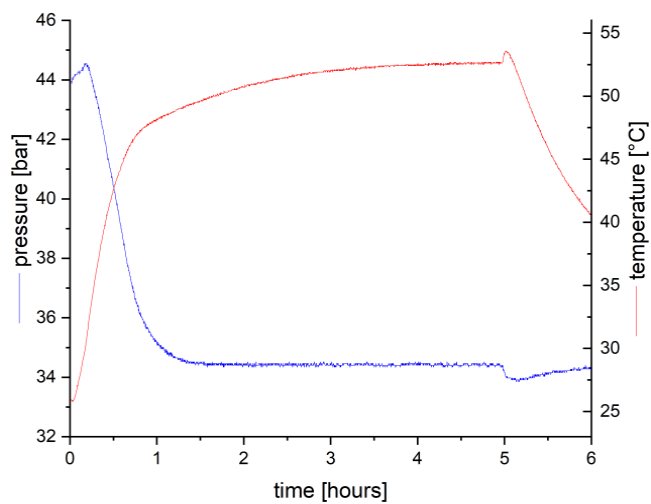

**Figure 1.** Pressure (blue line) and temperature (red line) changes during ball milling as a function of the milling time. The milling is stopped after 5 hours.

To obtain YHO, yttrium hydride produced via reactive milling inside a pressured hydrogen atmosphere needs to be oxidized, similarly to what was done to synthesize YHO after magnetron sputtering[2]. However, simple air exposure of nanoparticles, similar to how YHO thin films are synthesized, leads to a high oxidation rate due to a significantly larger surface area. Excess heat from the increased reaction rate results in a fully oxidized gray/white powder with no photochromic properties. This is similar to reports for YHO thin films at elevated temperatures where photochromic response lowered due to permanent oxidation [12] of unprotected thin films. The limited reports on variety of oxidation methods during the sputtering process[5], [6], [2], combined with the commonly reported YHO synthesis method of exposing sputtered yttrium hydride to air for oxidation by several research groups[13], [14], [15], [16], suggest that the presence of water in air has a significant impact on the oxidation reaction rate of the thin films. To limit the oxidation rate we placed the ball milled hydrogenated yttrium particles under dry air which is also containing lower oxygen percentage ($H_2O$ < 0.5 ppm, $O_2$ < 5-10%). Although a reduction in oxidation rate was not quantified, a clear difference in reactivity under normal and dry air observed. Under regular air, the ball-milled hydrogenated yttrium particles would rapidly ignite and fully oxidize with a visible reaction. In contrast, when exposed to dry air, the particles transitioned more slowly from a highly absorptive black color to a yellow color, as is typically observed in photochromic thin films[2]. The ball-milled particles were kept under the dry air atmosphere for up to two weeks to ensure the complete stabilization of the powder before exposure to regular air. After controlled oxidation, the powder was analyzed by SEM to evaluate the particle sizes. SEM images obtained at 5.0kV acceleration voltage are shown in Figure 2 with magnification from (a)100k to (d)3k. In Figure 2(a), particles with size below 500nm can be observed where presence of several sub-micron size particles in Figure 2(b-d) can also be observed. However, formation of smaller agglomerates on the larger particle (Fig.2 a) might suggest that improvement in ball milling process combined with a better oxidation control can result in a further particle size reduction.

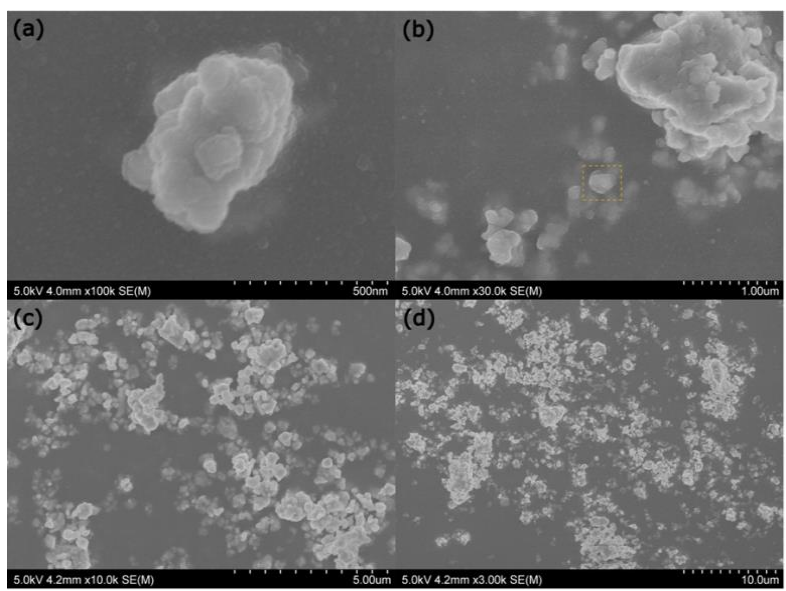

**Figure 2.** SEM images of the YHO powder after controlled oxidation shown at magnification of (a)100k, (b)30k, (c)10k and (d)3k.

Photochromism in YHO is typically investigated through changes in transmittance, however, the opaque nature of these powders in this study required an investigation of their optical properties through reflectance measurements. The optical properties of ball milled powder after controlled oxidation are shown in Figure 3(a and b). To investigate the photochromic behaviour, reflectance data was obtained by placing the powder sample between two boro-silicate glass and measured using an integrated sphere. Photo-darkening was achieved by exposing the sample to the light from a 405 nm peak purple light at an irradiance of 66 mW/cm2 for 90 minutes (see blue curve in Figure 3[a] as compared to the red curve for the initial state). Photochromic contrast level, shown through change in reflectance, was around 10% at 850nm, showing reversible photochromatic properties where the majority of the contrast was recovered after 5 minutes under dark (yellow curve in Figure 3[a]). Further, the cyclic photochromic switching behavior was investigated by recording and extracting the red component from images and capturing the cyclic change between the darkened and bleached states (Figure 3[b]). The relative contrast level was calculated by $(R_0-R)/R_0$, where R is the signal value measured from the red component from the images and $R_0$ is the initial value at clear state. Photochromic sample powder was cycled for 9 cycles consisting of 6 minutes periods of illumination and dark. Initially, sample showed around 10% contrast while a clear state was not fully achieved at the beginning of each cycle, consistent with the well-reported behavior of YHO thin film referred to as 'memory-effect' [17], [3]. After 9 cycles, the darkest contrast relative to the initial value was measured around 30%. The result shows that the YHO powders produced via ball milling demonstrate photochromic properties comparable to thin films [12], [18].

To confirm the presence of YHO phase within the powders obtained through the reactive ball milling and controlled oxidation, X-ray diffraction analysis was performed. The resulting pattern is shown in Figure 3(c), which compares the position of the diffraction peaks with the values reported for YH2 (JCPDS - 04-006-6935). The experimentally observed diffraction peaks are shifted to lower angles consistent with an expanded lattice parameter due to the incorporation of oxygen, confirming the formation of YHO [19], [2].

The YHO powder was mixed with a polystyrene (PS) solution and cast onto a glass substrate. After the coating dried, a mask was applied and the sample illuminated. This process aimed to spatially control the photochromic effect, darkening exposed areas while masked regions remained unchanged. Figure 3(d) shows the resulting pattern, highlighting the contrast between the illuminated and masked regions. The darker areas correspond to the portions exposed to light, demonstrating photochromic darkening of the YHO-PS composite. In contrast, the lighter masked areas, retained their original color, visually confirming the localized nature of the photochromic response. This technique demonstrates the potential for using YHO-PS composites for applications requiring rewritable patterns or images, similar to focal plane masks described in [20]. Essentially, the YHO-PS composite acts as a rewritable medium, where patterns can be written (photo-darkened) and erased (bleached) using light, offering a dynamic method for creating and modifying spatial information. However, the initial contrast is lower than that observed in thin films. Therefore, further optimization of the synthesis is needed to enhance performance and unlock new application possibilities.

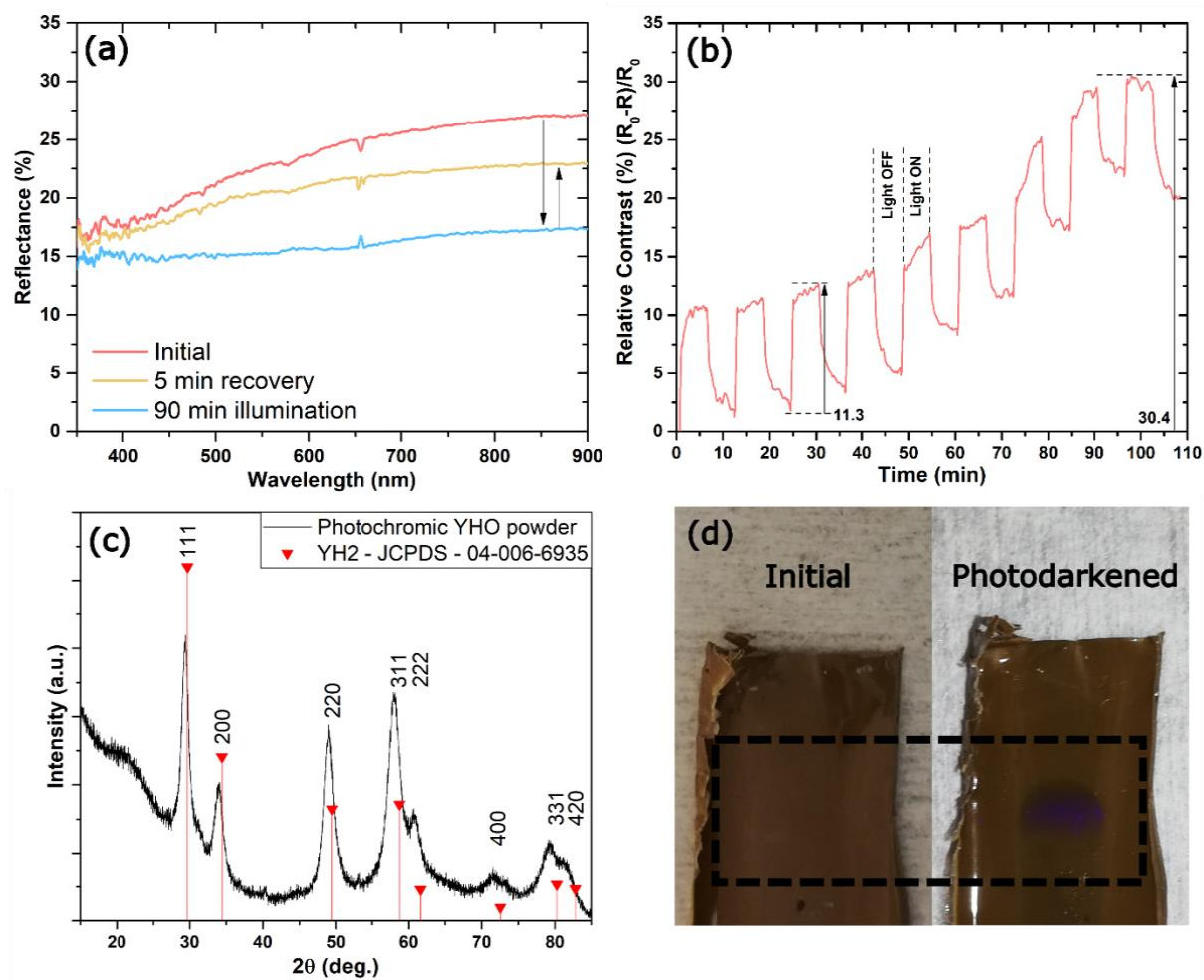

**Figure 3.** Photochromic properties of YHO powder were demonstrated through photodarkening and bleaching by UV illumination. (a)Reflectance and (b) contrast ($[R_0-R]/R_0$) values demonstrates the optical modulation before illumination (initial-red circle), after illumination (90 minutes – purple down triangle) and after storing in darkness (5 minutes bleaching – blue up triangle). (c) Powder X-ray diffraction pattern showing characteristic YHO phase with peak shifts to lower angles compared to $YH_2$ reference (JCPDS - 04-006-6935), confirming oxygen incorporation. (d) Spatially controlled photochromic patterning demonstrated by selective UV illumination of YHO-polystyrene composite film through a mask, showing darkened exposed regions and unchanged masked areas.

This study demonstrates the successful synthesis of YHO powders exhibiting photochromic properties via reactive ball milling and controlled oxidation. The resulting materials exhibit optical behaviors, including darkening under 405 nm illumination, bleaching in darkness, and reversible cyclic switching, that are consistent with previously reported characteristics reported for YHO thin films [21], [15]. Although the initial contrast is lower than that of thin films, the ability to achieve photochromic switching, especially the spatially controlled patterning achieved using the PS composite, confirms the viability of this powder-based approach. Importantly, this approach enables scalable production of photochromic YHO powders. In contrast to thin film deposition techniques, which face limitations in throughput and scalability, ball milling is inherently suited for large-scale processing [22]. This advantage positions YHO powders as promising candidates for industrial applications requiring photochromic functionality potentially expanding their use across a broader range of technologies.

## Conclusion

This work demonstrates the feasibility of producing photochromic YHO powders via reactive ball milling and controlled oxidation. The resulting material exhibits reversible photochromic switching under 405 nm illumination, aligning with the general behavior observed in YHO thin films. While the photochromic contrast in the powders is initially lower than in thin films, the scalable nature of ball milling presents a significant advantage for bulk production. This scalability is crucial for potential industrial applications, contrasting with the limitations of thin film deposition techniques. Further research focusing on optimizing milling parameters, oxidation control, and incorporating stabilizing matrices, could enhance the photochromic contrast and enable to a wider range of applications, from smart windows and displays to rewritable media. The inherent scalability of ball milling, combined with the flexibility of the synthesis route, positions YHO powders as a promising alternative to thin films for large-scale deployment of photochromic technologies.